\documentclass[]{tGIS2e}
\usepackage{comment}
\usepackage{graphicx}
\usepackage{amsmath}
\usepackage{amssymb}
\usepackage{eurosym}
\usepackage[table,xcdraw]{xcolor}
\definecolor{green}{rgb}{0,0.8,0}
\usepackage{algorithm,algpseudocode}
\usepackage{caption}
\usepackage{url}
\usepackage{framed} 
\usepackage{lineno}

\begin{document}



\title{Estimating the Prediction Performance of Spatial Models via Spatial k-Fold Cross Validation}

\author{Jonne Pohjankukka$^{\ast}$\thanks{$^\ast$Corresponding %
author email: jjepoh@utu.fi \vspace{6pt}}, Tapio Pahikkala, Paavo Nevalainen and Jukka Heikkonen\\\vspace{6pt} {\em{Department of Future Technologies, University of Turku\\ Vesilinnantie 5, 20500 Turku.}}\\ \vspace{6pt}}


\received{released July 2017}

\maketitle

\begin{abstract}
In machine learning one often assumes the data are independent when evaluating model performance. However, this rarely holds in practise. Geographic information data sets are an example where the data points have stronger dependencies among each other the closer they are geographically. This phenomenon known as spatial autocorrelation (SAC) causes the standard cross validation (CV) methods to produce optimistically biased prediction performance estimates for spatial models, which can result in increased costs and accidents in practical applications. To overcome this problem we propose a modified version of the CV method called \textit{spatial k-fold cross validation} (SKCV), which provides a useful estimate for model prediction performance without optimistic bias due to SAC. We test SKCV with three real world cases involving open natural data showing  that the estimates produced by the ordinary CV are up to 40\% more optimistic than those of SKCV. Both regression and classification cases are considered in our experiments. In addition, we will show how the SKCV method can be applied as a criterion for selecting data sampling density for new research area. \bigskip

\bigskip

\end{abstract}
\setlength{\textfloatsep}{-10pt}
\section{Introduction}
\begin{figure}[!htbp]
	\centering
  	\includegraphics[width=0.8\textwidth,height=6cm]{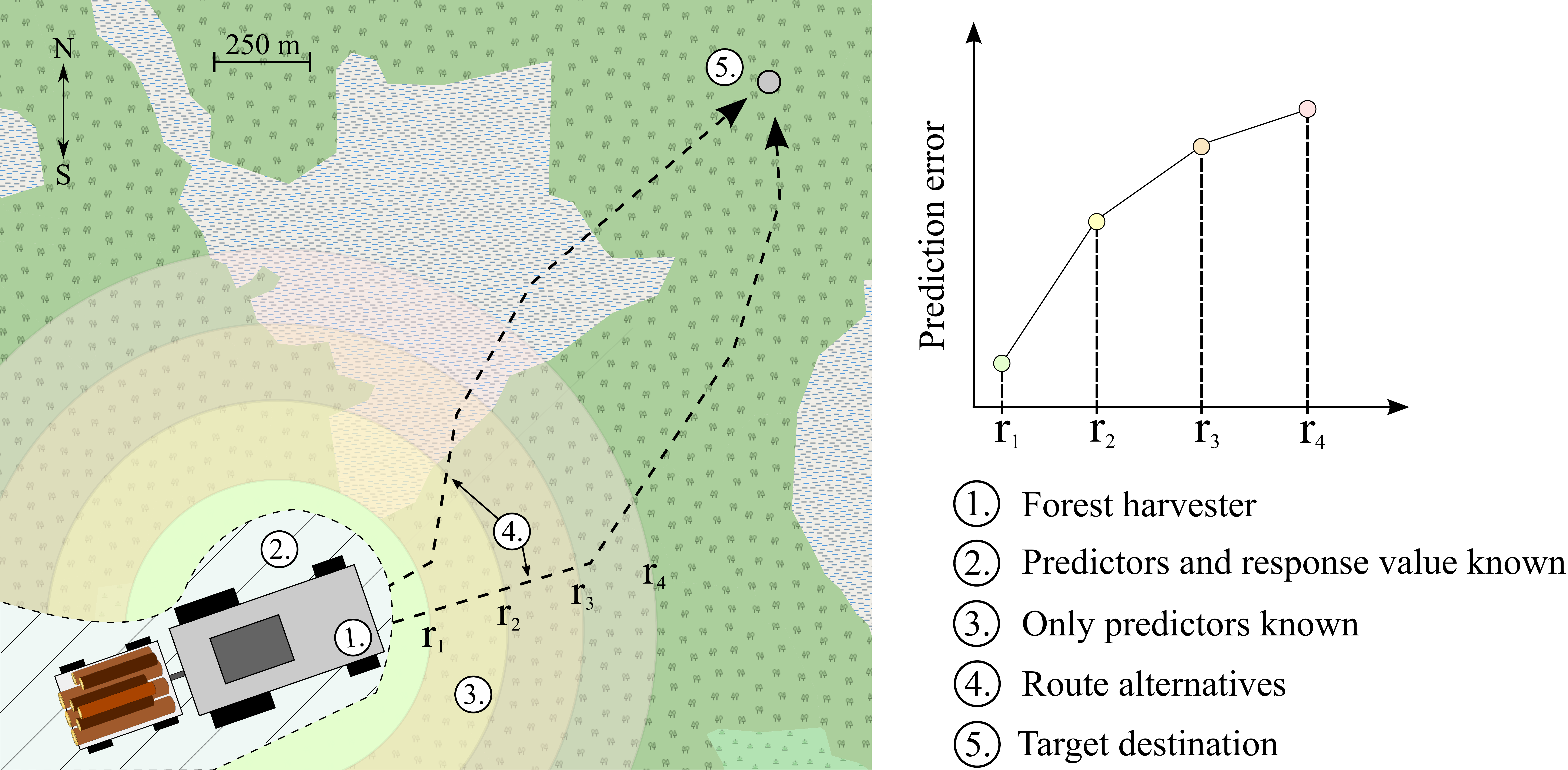}
    \vspace{2mm}
    \caption{The forest harvesting example. The harvester driver needs to select an optimal route to target destination. Due to SAC it is to be expected that the prediction error increases the further away we make point predictions. The background map in the image (also in Figure \ref{fig:3by3grid}) is by the courtesy of OpenStreetMap.}
    \label{Fig:harvester_illustration} 
\end{figure}

An important step in machine learning applications is the evaluation of the prediction performance of a model in the task under consideration. For this one can use the k-fold cross validation (CV), which assumes the data are independent. Geographic information system (GIS) data sets represent an example where the independence assumption naturally does not hold due to the temporal or spatial autocorrelation (SAC). SAC and its effects on spatial data analysis has been extensively studied in spatial statistics literature, e.g. \citep{Legendre1993,Koenig1999}. For example it has been shown that the failure to not account the effect of SAC in spatial data modeling can lead to over-complex model selection \citep{Hoeting2006,Rest2014}. Generally speaking, natural data exhibits SAC because of the first law of geography and fundamental principle in geostatistical analysis according to Waldo Tobler \citep{Tobler1970}: "Everything is related to everything else, but near things are more related than distant things". In spatial statistics the degree of SAC of a data set can be measured using e.g. a semivariogram \citep{Variogram2015}, Moran's I \citep{Moran50}, Geary's C \citep{Geary54} or Getis's G \citep{Getis92}.\\ \indent  There are numerous applications involving spatial data which have problems caused by SAC in the data sets such as, natural resource detection, route selection, construction placement, natural disaster recognition, tree species detection, environmental monitoring etc. \citep{AlaIlomaki2015}. Consider the example of harvesting operations in forestry where optimal route selections are of key importance. In order to minimize the risk of harvester sinking into the soil a route with the optimal carrying capacity is required. The route selection is based on predictions of soil types along the route which gives the harvester an estimate on the carrying capacity of the route. If the effect of SAC is not considered in the soil type predictions while estimating the model performance we might end up selecting a hazardous route. The reason for this is that the spatial model we are using gives over-optimistic prediction performance for soil types farther away from the harvester's current location. The model implicitly assumes that we have known soil types close to the predicted soil types which is not always the case. This fact must be taken into account in the model prediction performance evaluation in order to avoid over-optimistic estimation. An illustration of the considered example is shown in Figure \ref{Fig:harvester_illustration}.


\vspace{0.5cm}

\indent To counter the problems caused by SAC in spatial modeling one usually tries to incorporate SAC as an autocovariate factor into the prediction models themselves, e.g. autocovariate models, spatial eigenvector mapping, autoregressive models \citep{Brenning2005,Dormann2007a,Bahn2006,Betts2009,Beale2010,Lichstein2002,Diniz-Filho2003,Zhang2010}. A review of such methods is well presented in \citep{Dormann2007a}. Other methods include spatial clustering and re-sampling techniques for countering SAC \citep{Hijmans2012,RuBKruse2010,Brenning2012}. Despite the vast literature of techniques for spatial prediction little attention is given for assessing the spatial prediction performance of a model via cross validation techniques. In \citep{Cressie2015} the author does not advocate CV for confirmatory data analysis because the independence assumption in the data samples is inherently not valid in geostatistical context.

In this article, we propose a novel CV method called \textit{spatial k-fold cross validation} (SKCV) for estimating prediction performance under SAC-based independence violations in the data. SKCV is also applicable for selecting grid sampling density for new research areas. More specifically, SKCV attempts to answer the following two questions:

\begin{framed}
\begin{enumerate}
\item What is the prediction performance of a model at a certain geographical location when the closest data measurements used to train it lie at a given geographical distance? 
\item Conversely, if the prediction performance is required to be at least a given level, how dense data sampling grid should be used in the experiment area to achieve it?
\end{enumerate}
\end{framed}

\noindent The question (2) is about the trade-off between the prediction performance and data collection costs. The SKCV method provides the model prediction performance as the function of geographical distance between the in-sample and out-of-sample data, and hence it indicates how close geographically  training data has to be to the prediction area in order to achieve a required prediction performance. The idea in SKCV is to remove the optimistic bias due to SAC by omitting data samples from the training set, which are geographically too close to the test data.

To evaluate how well SKCV answers the above questions, it is tested with three real world applications using public GIS-based data sets. The applications involve assessing the predictability of water permeability of soil and forest harvest track damage. Both regression and classification models were used in these experiments. The usability of the SKCV method for determining the needed sampling grid density is tested by measuring the difference between the performance of model constructed with a given grid density and the result predicted with SKCV. We will explain this comparison in more details in section 4.1.

We wish to emphasize that we use SKCV in this manuscript for assessing the spatial prediction performance of a model and not for model complexity selection even though model complexity selection can also be applied with SKCV. In \citep{Rest2014} the authors used a similar spatial cross validation method as SKCV for model variable selection. In their work, they compared a special case of SKCV method, the spatial leave-one-out method (SLOO) with Akaike information criterion (AIC) \citep{Akaike73} as a criterion for model variable selection. It turned out that SAC caused the AIC to select biased variables, whereas SLOO prevented this. In \citep{Pohjankukka2014b,Pohjankukka2014a,Pohjankukka2015} the SKCV method was called \textit{cross validation with a dead zone} method. Related studies on spatial data analysis can also be found in the works of \citep{Azz94,Brenning2012,Schu05}.\\ \indent In what follows, a formal description of the SKCV method will be given in Ch. 2, followed by description of used data sets in Ch. 3 and experimental analyses with three sample cases in Ch. 4, and finally Ch. 5 includes conclusions.

\section{Spatial k-fold cross validation}
\begin{figure}[!htbp]
	\vspace{38pt}
	\centering
  	\includegraphics[scale=0.5]{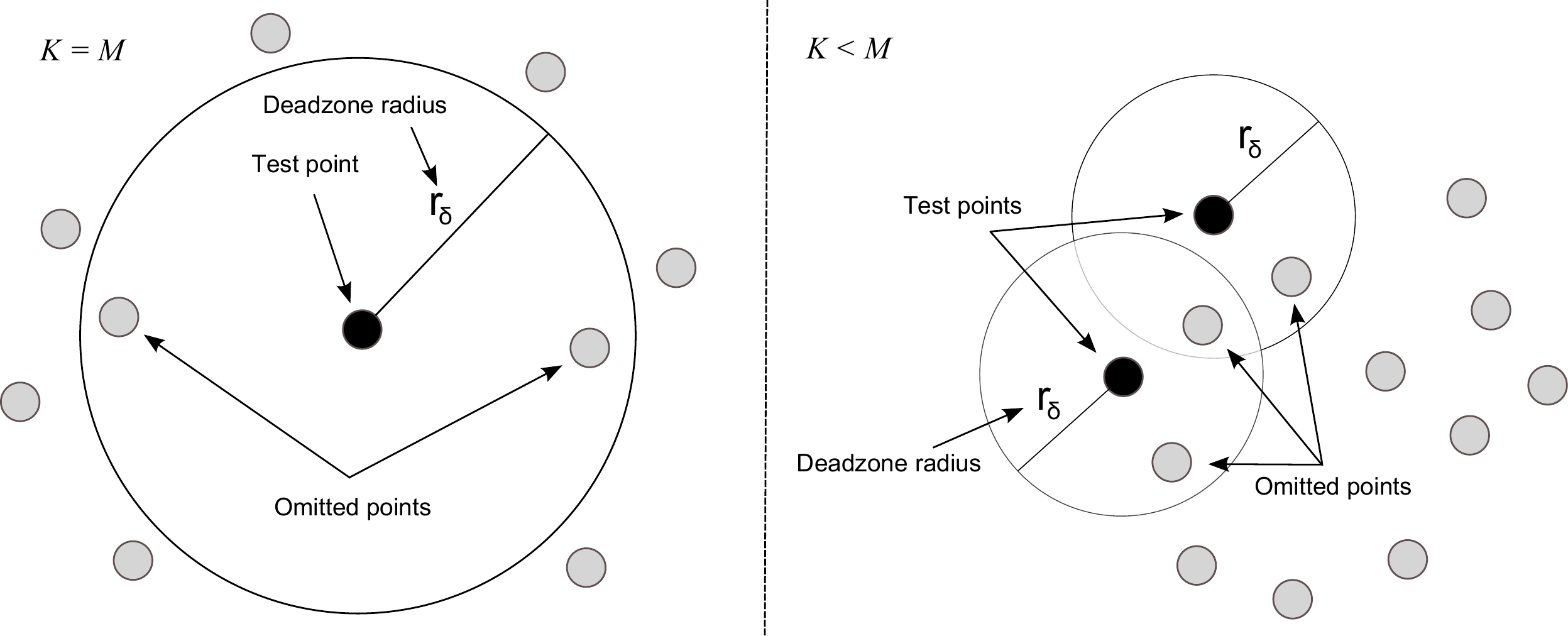}
    \vspace{2mm}
    \caption{Reduction of the training set in the SKCV procedure. The black and gray points correspond to test and training data points. The gray data points inside the perimeters of radius $r_\delta$ are omitted from the training data, after which the test points are predicted using the remaining training data (i.e. the gray data points outside the perimeters).}
    \label{Fig:SKCV_illustration}
\end{figure}

SKCV is a modification of the standard CV to overcome the biased prediction performance estimates of the model due to SAC of the data. The over-optimistic bias in the performance estimates is prevented by making sure that the training data set only contains data points that are \textit{at least} a certain spatial or temporal distance away from the test data set. \\ \indent We will denote our data point as $\textbf{d}_i = (\textbf{x}_i, y_i, \textbf{c}_i)$, where $\textbf{x}_i\in \mathbb{R}^n$ is a feature vector, $y_i\in \mathbb{R}$ a response value and $\textbf{c}_i\in \mathbb{R}^2$ the geographical coordinate vector of $i$th data point. The data set is denoted as $\mathcal{D}=\{\textbf{d}_1, \textbf{d}_2, ..., \textbf{d}_M\}$. The value $r_\delta\in\mathbb{R}^+$ is the so-called \textit{dead zone} radius, which determines the data points to be eliminated from the training data set at each SKCV iteration. The set $\mathcal{V}=\{\mathcal{V}_1, ..., \mathcal{V}_K\}$ is the set of cross validation folds, where each $\mathcal{V}_p\subset\mathcal{D}$ and $\mathcal{V}_p \cap \mathcal{V}_q = \emptyset$, when $p\neq q$ and $\bigcup_{p=1}^K \mathcal{V}_p = \mathcal{D}$. The training of the model is performed by a learning algorithm $\mathcal{A}$. The vector $\hat{\textbf{y}}\in \mathbb{R}^M$ denotes the predicted response values by a prediction model $\mathcal{F}$. Note that the choice of $\mathcal{F}$ does not affect the functionality of SKCV. We use the standard Euclidean distance $e$ to calculate the \emph{spatial distance} between two data points $\textbf{d}_i$ and $\textbf{d}_j$. A formal presentation of the SKCV method is given in Algorithm \ref{Algo}. When the number of folds $K$ equals the number of data points $M$ SKCV becomes SLOO. The SKCV algorithm is almost identical to normal CV with the exception of the reduction of the training set depicted in Figure \ref{Fig:SKCV_illustration} and in line 2 of Algorithm \ref{Algo}. In particular, when $r_\delta = 0$ SKCV reduces to normal CV. 

\begin{algorithm}
  \caption{Spatial k-fold cross validation}\label{Algo}
  \begin{algorithmic}[1]
    \Require $\mathcal{V}, \mathcal{D}, \mathcal{A}, r_\delta$
    \Ensure $\hat{\textbf{y}}$
      \For{$i \leftarrow 1$ to $K$}
        \State $\mathcal{H} \leftarrow \bigcup_{\textbf{d}_k \in \mathcal{V}_i} \left\{\textbf{d}_j\in \mathcal{D}\;|\; e(\textbf{c}_j, \textbf{c}_k) \leq r_\delta \right\}$\Comment{Remove data points too close}
        \State $\mathcal{F} \leftarrow \mathcal{A}\left(\mathcal{D}\setminus\mathcal{H}\right)$\Comment{Build model using reduced training set}
        \For{$\textbf{d}_k \in \mathcal{V}_i$}
              \State $\hat{\textbf{y}}[k] \leftarrow\mathcal{F}\left(\textbf{x}_k, \textbf{c}_k\right)$\Comment{Make prediction}
         \EndFor
      \EndFor
      \\
      \Return $\hat{\textbf{y}}$\Comment{The predicted $\hat{\textbf{y}}$}
  \end{algorithmic}
\end{algorithm}


\indent There are three issues one might consider with SKCV which we will address here. Firstly, since SKCV may involve removal of a large number of training data, this may introduce an extra pessimistic bias on the prediction performance not related to SAC. The size of this bias can be estimated via experiment in which one removes the same amount of randomly selected data from the training set on each CV round. Our experimental results in later sections confirm that the performance decrease observed by doing this is negligible compared to the one caused by SAC removal.

Secondly, the above considered issue becomes far more severe if the number of SKCV folds $K$ is very small (say $K=2$). It could happen that most of the training data is removed because the combined dead zones of the test data points will have a large \emph{effective radius}. This concern is application specific and the selection of the SKCV folds must be designed to suit the purposes of the application. For example with a sparse data set it would make a little practical sense to select the SKCV folds in such a way that all the training data is removed. For these reasons it is best to have $K=M$, which corresponds to the SLOO case of SKCV if computational resources allow it.

Thirdly, one could ask whether the prediction performance for a given $r_\delta$ could be estimated by analyzing the prediction error obtained with, say, leave-one-out cross-validation simply as a function of the average distance to closest neighbors. While, this could be doable with data sets having both densely and scarcely measured areas, the data points in many available data sets tend to be much closer to each other than in the case we intend to simulate. For example, with a dense data set with a maximum distance of $3$ m between a data point and its nearest neighbors, one can not simulate performing prediction for a data point having the closest measurements at least 25 meters away.

Finally, let us consider the difference between spatial interpolation and regression. In the former, the only extra information available about the training data are their coordinates $\textbf{c}$, while in the latter one also has access to an additional information in the form of feature representation $\textbf{x}$. However, the SKCV algorithm works in a similar way in both cases, as it is independent on the type of information the learning algorithms use for training a model or what the model uses for predicting the responses for new points.


\begin{figure}[b]
\vspace{38pt}
\begin{center}
\resizebox{\textwidth}{!}{%
\begin{minipage}{150mm}
\subfigure[]{
\resizebox*{4.8cm}{!}{\includegraphics{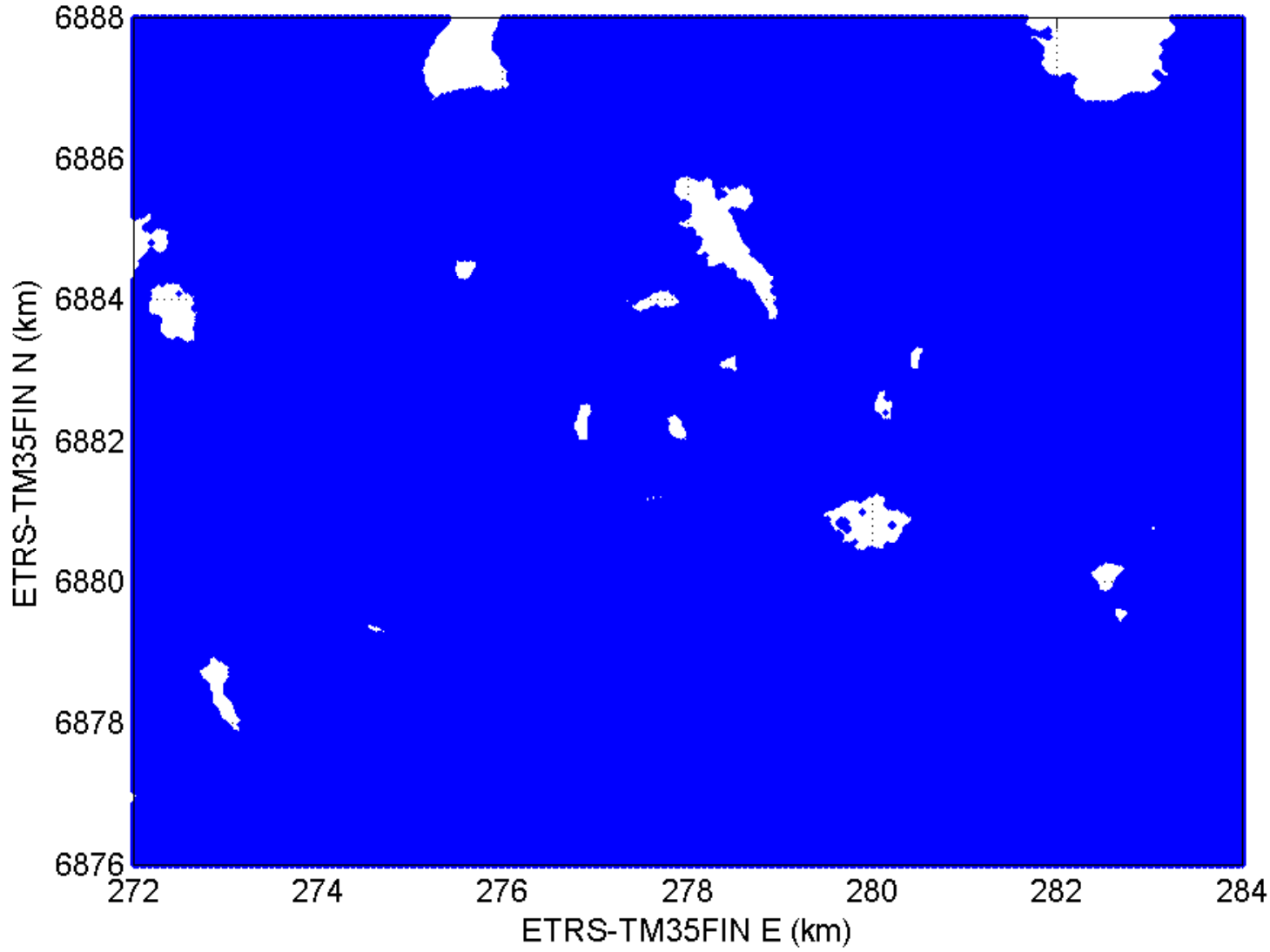}}}%
\subfigure[]{
\resizebox*{4.8cm}{!}{\includegraphics{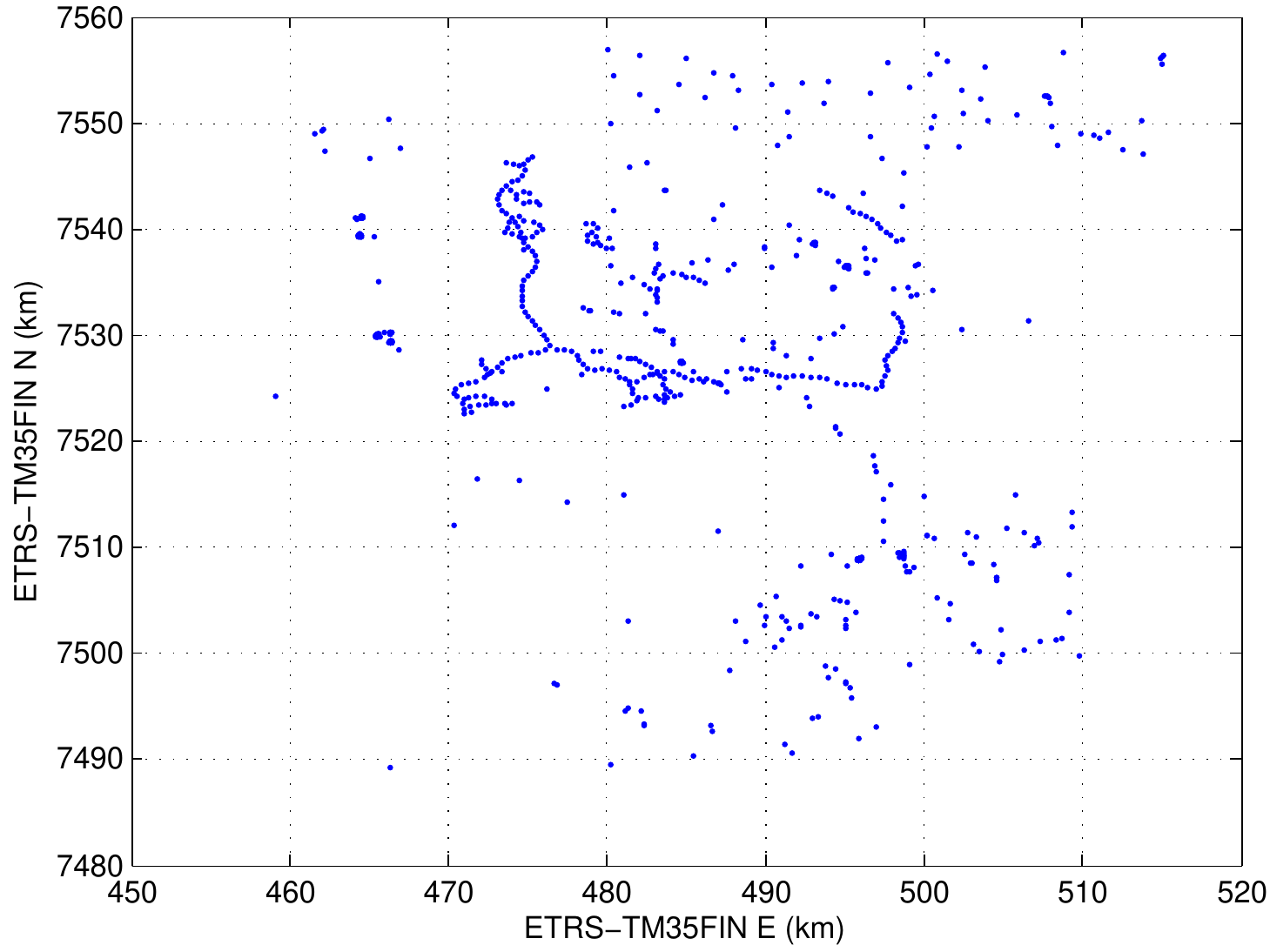}}}%
\subfigure[]{
\resizebox*{4.8cm}{!}{\includegraphics{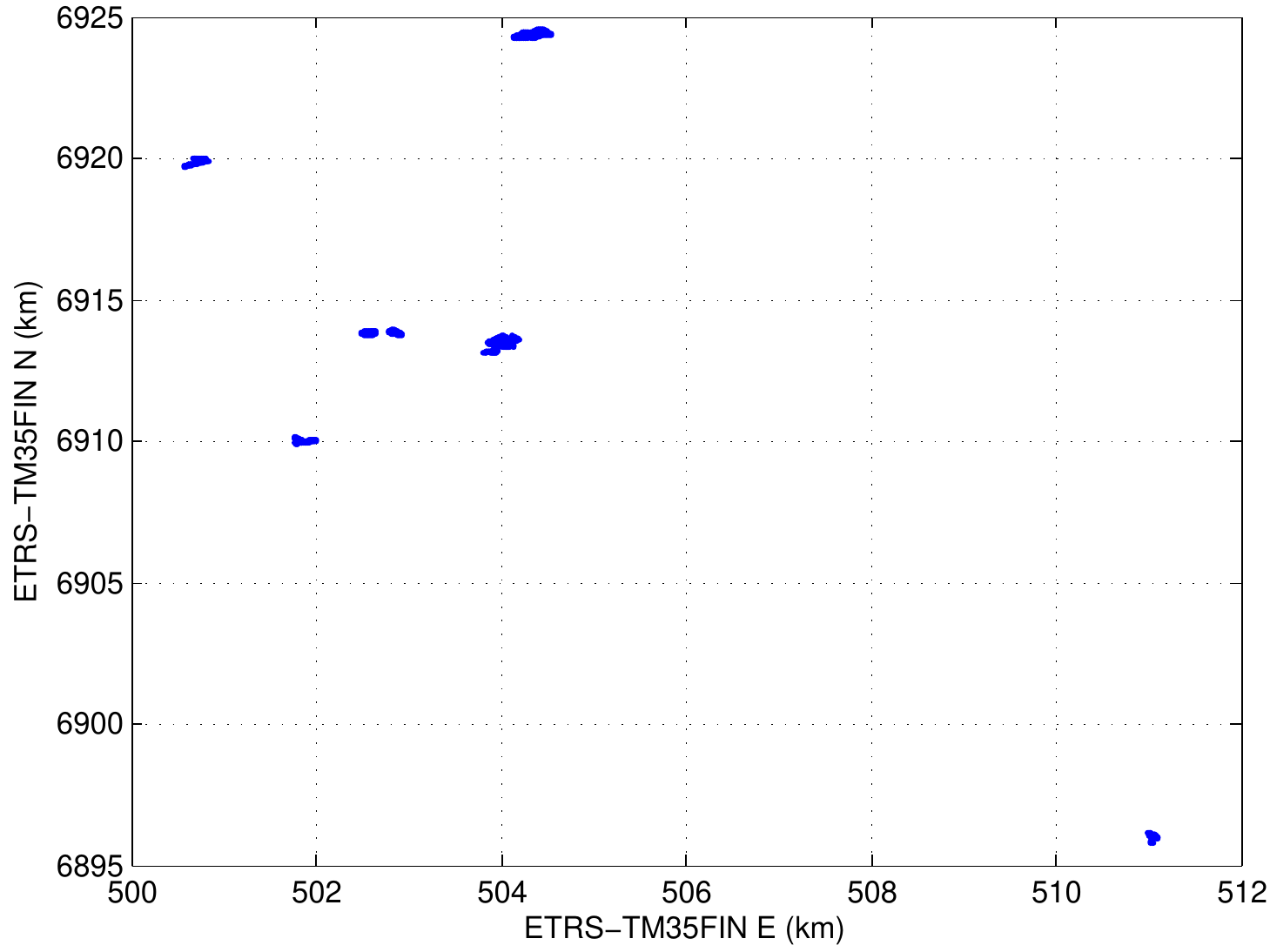}}}%
\caption{Coverage of data on experimental cases 1-3. (a) case 1: 361201 data points, (b) case 2: 1691 data points, (c) case 3: 11795 data points. Blue areas correspond to areas where data was available. The axes correspond to locations in ETRS-TM35FIN coordinates in kilometers. The data set in case 1 is much more dense than data sets in cases 2 and 3.}%
\label{fig:DATA_AVAILABILITIES}
\end{minipage}
}
\end{center}
\end{figure}

\section{Data sets}
The three experimental cases differ on the availability and resolution of the data sets. In Case 1 related to water permeability prediction data was available throughout the research area, with the exception of areas where there were obstacles (e.g. buildings or lakes). In case 2 also related to water permeability prediction, there were scattered field measurement data and in case 3 related to harvester track damage prediction the data set was clustered into several areas. These cases are typical of common types of spatial prediction applications. The availability of the data sets in three cases is illustrated in Figure \ref{fig:DATA_AVAILABILITIES}. The data range from remote sensing data sets such as satellite and airborne imaging raster data to manually on-site collected samples of the soil \citep{Pohjankukka2014b,Pohjankukka2014a,Pohjankukka2015,Wood1996,Zeven1987,Tomppo2008,Hyvonen2003}. The formats of the data sets are TIFF-images and ASCII-files with different resolutions. A summary of the used data sets in the three cases is illustrated in Table \ref{table:case_data_sets}. In the following paragraphs we briefly describe the used data sets. More detailed illustration of the data sets is given in the supplementary material.

\textbf{Digital elevation model data}:
we downloaded digital elevation model (DEM) data
from the file service for open data by the National Land
Survey of Finland (NLS). The DEM was made from airborne
laser scanning data with grid size of 2 m. Several geomorphometric variables were
derived from the NLS DEM in SAGA GIS environment.
In our analysis we used the geomorphometric features:
plan curvature, profile curvature, slope, topographic wetness
index, flow area, aspect, diffuse insolation and direct
insolation \citep{Zeven1987,Wood2009,Wood1996,Beven1979,Seibert2007}. These derived features are more efficient for prediction than raw height data alone.

\textbf{Multi-source national forest inventory data}: a selected set of 43 features of the state of the Finnish forests in 20 m grid size are available as the Multi-Source National Forest Inventory (MS-NFI) by Natural Resources Institute Finland (LUKE). The MS-NFI data set is derived by interpolating field measured MS-NFI samples using inverse distance-weighted k-nearest neighbor method as the interpolation algorithm and Landsat imagery combined with DEM data as the basis of interpolation. Features include e.g. the biomass and volume of growing stock. The MS-NFI data features exhibit built-in dependencies which means the number of useful features is lower than 43. A detailed description of the MS-NFI is \citep{Tomppo2008}.

\begin{table}[t]
	\vspace{38pt}
	\footnotesize
	\begin{center}
    \caption{Summary of the used data sets in all experimental cases. Response value data sets are listed in emphasized form. Also the data format is shown either as TIFF-raster image or ASCII-vector file and grid resolution size in meters.}
	\label{table:case_data_sets}	
	\vspace*{5mm}
    	\resizebox{\textwidth}{!}{%
		\begin{tabular}{| l | c | c | c | c | c |}
 			\hline
 			\cellcolor{gray!25}\textbf{Data set} & \cellcolor{gray!25} \textbf{Format} & \cellcolor{gray!25} \textbf{Grid size} &  \cellcolor{gray!25} \textbf{Case 1} & \cellcolor{gray!25} \textbf{Case 2} & \cellcolor{gray!25} \textbf{Case 3}\\\hline
 			Digital Elevation Model & Raster & 2m & X & X & X \\ \hline
  		    Multi-source National Forest Inventory & Raster & 20m & X & X & X  \\ \hline
  		    Gamma-ray spectroscopy & Raster & 50m & X & X & X  \\ \hline
  		    Air-borne electromagnetic & Raster & 50m &  & X &   \\ \hline
            Peatland & Raster & 1000m &  & X & X   \\ \hline
  		    Weather & Raster & 10000m &  &  & X  \\ \hline
  		    Stoniness &  Vector & -  & &  & X  \\ \hline
  		    Soil moisture & Vector & -  &  & & X  \\ \hline
  		    \textbf{Water permeability exponent} & Vector & - & X & X &   \\ \hline
  		    \textbf{Harvester track damage} & Vector & - &  &  & X  \\ \hline
		\end{tabular}
        }
	\end{center} 
\end{table}

\indent \textbf{Aerial gamma-ray spectroscopy data}:
the aerial gamma-ray flux of potassium (K) decay with the grid size of 50 m is provided by the Geological Survey of Finland (GTK). This data is related to e.g. the moisture dynamics, frost heaving \citep{Hyvonen2003} and density, porosity and grain size of the soil. High gamma-radiation indicates lower soil moisture and vice versa. Several statistical and textural features were derived from the gamma-ray data. These include: 3x3 windowed mean and standard deviation, Gabor filter features \citep{Feichtinger} and Local Binary Pattern features \citep{Pietikainen}.

\textbf{Peatland data}:
the peatland data is provided by LUKE and uses topographic information provided by 
NLS. The peatland data is a binary raster mask of 1000 m grid size with values 0/1 corresponding to non-peatland/peatland areas. The peatland mask is derived from four NLS topographic database elements depicting different type of peatlands.
The mask bit 1 refers to a spot where the location is mostly covered by peatland 
vegetation and the peat thickness exceeds 0.3 m over an local area of 1000 m$^2$.

\textbf{Air-borne electromagnetic data}:
the air-borne electromagnetic (AEM) data was provided by the GTK. The apparent resistivity indicates the soil type factors, e.g. grain size distribution,
water content and quality in the soil and cumulative weathering.

\indent \textbf{Weather data}:
weather data on temperature (C) and rainfall (mm) for years 2011-2013 was
provided by the Finnish Meteorological Institute (FMI). The grid size of the data set was 10 km. We used the mean temperature and rainfall of the last 30 days 
at each observation point of the response value.

\textbf{Stoniness data}:
stoniness was estimated by steel-rod sounding \citep{Tamminen1991}. The rod was pushed into the soil where the penetration depth and stone hits were recorded.

\textbf{Soil moisture data}:
gravimetric soil water content was measured from the samples by drying the soil samples and calculating the weight difference of dry and wet soil sample \citep{ASTM2}.

\textbf{Water permeability exponent data}:
water permeability indicates the nominal vertical speed of water through the soil sample. This feature was measured indirectly by observing the soil particle size distribution. The actual speed depends on inhomogenities (roots, rocks) and micro-cracks in the soil. The water permeability exponent is a logarithmic quantity $y$ derived from water permeability speed $v$.

\textbf{Harvester track damage data}:
approximately 36 km of strip roads were traversed by Mets\"ateho Ltd. and visually assessed into damage classes by a forest operations expert. The soil damage classes used were: (1) No damage; (2) Slight damage; and (3) Damage. The original dataset required preprocessing by LUKE due to the inaccuracies in GPS-tracks. The strip road line segments were then converted to sample points used in the prediction process.

\section{Experimental analysis with SKCV} 
In this section the SKCV method is applied to three real world cases involving GIS-data making them suitable to illustrate the proposed method. In the first two cases the water permeability levels of boreal soil are predicted and in the final case the damage caused by movements of a forest harvester. The experiments provide useful results e.g. for forest industry where it is crucial to have accurate and optimistically unbiased prediction performance for soil conditions. It is estimated that forest industry in Finland alone has yearly costs of approximately 100M\euro\, caused by challenging trafficability conditions of the soil which increase time and fuel consumption and decrease the efficiency of timber harvesting operations \citep{Pennanen2003,Siren2013}. These costs could be decreased by additional information on soil conditions, especially soil bearing capacity by utilizing public GIS-data.

The research question (1) will be addressed in cases 1, 2, 3 (sections 4.1, 4.2, 4.3) and the research question (2) will be addressed in case 1 (section 4.1). In all experimental cases $k$-nearest neighbor (kNN) algorithm was used as the prediction model $\mathcal{F}$ and the predictor features $\textbf{x}_i$ were z-score standardized. While there are many alternative prediction methods the choice does not have an effect on the presence of SAC in the data, therefore kNN was selected due to its simplicity. As a distance function that determines the nearest neighbors we use the Euclidean distance for the feature vectors $\textbf{x}_i$. Note that this is in contrast to the spatial distance $e$ used in SKCV.
We implemented the analyses using $k$-values of $\{1,3,5,7,9,11,13,15\}$ for kNN. The general behavior of SKCV results was similar for all tested $k$-values and for this reason we only report the results with $k=9$. The performance measures used in the experiments were the standard root mean squared error (RMSE) for kNN-regression \citep[p. 66--73]{kNNREG2014} and classification accuracy for kNN-classification. In cases 1 and 2 (regression) the predicted response value $\hat{y}_i$ is defined as the average value of k-nearest neighbors and in case 3 (classification) the mode of the k-nearest neighbors.


\begin{figure}[b]
	\vspace{38pt}
	\centering
    \includegraphics[scale=0.25]{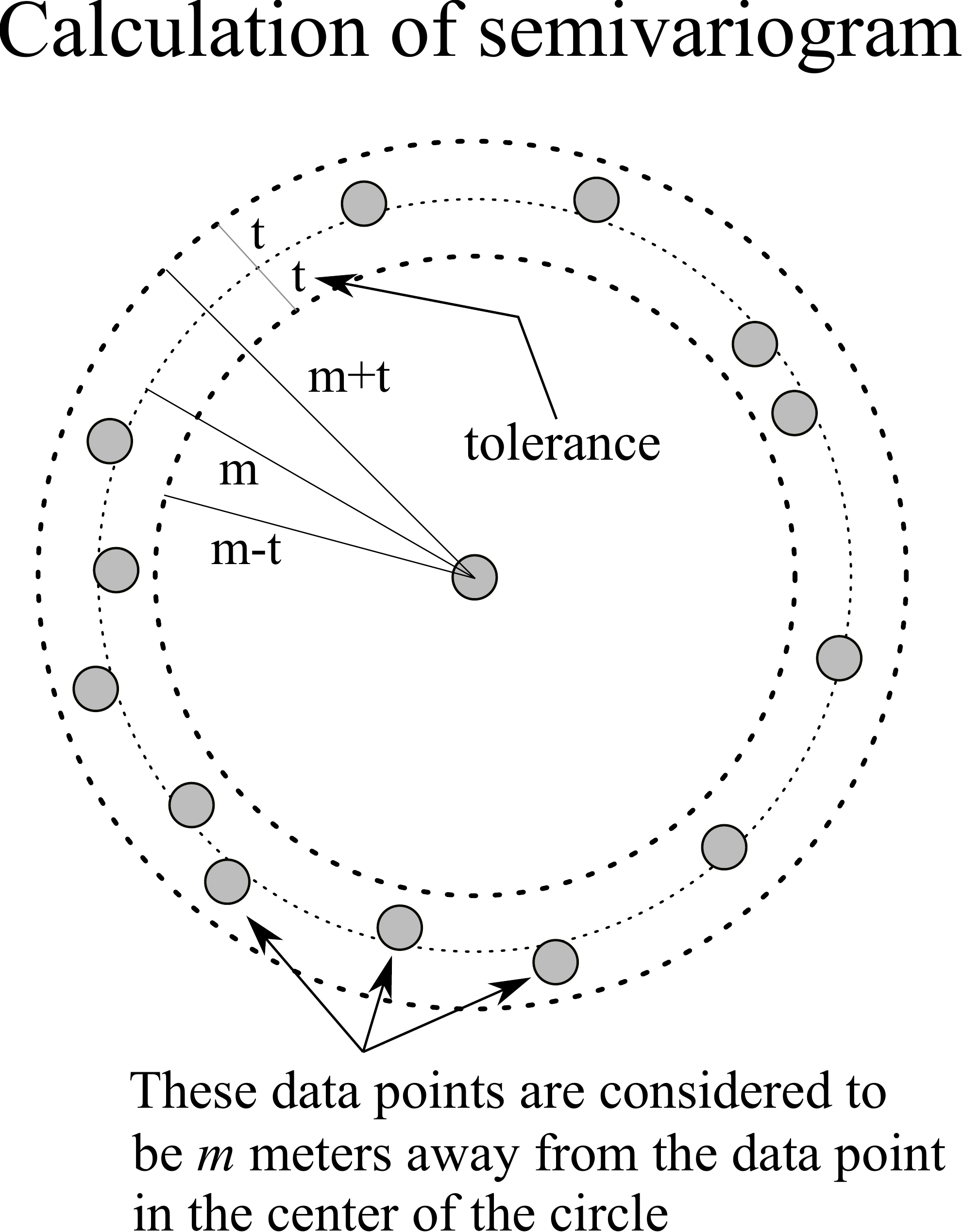}
    \caption{Calculation of the semivariogram. Data points within a distance range $[m-t, m+t]$ are considered to be $m$ meters away from the center data point.}
    \label{Fig:semivariogram}
\end{figure}

The semivariograms and Moran's I statistics were calculated for the response variables $y_i$ in all experimental cases to confirm the presence of SAC in the data. In a semivariogram, a variable $X$ is spatially autocorrelated at a given distance range $[m-t, m+t]\subset \mathbb{R}^+$ with \emph{lag tolerance} $t\in \mathbb{R}^+$ if its semivariogram value $\gamma_t(m)\in \mathbb{R}^+$ is lower than the \textit{sill value} of the variable $X$ \citep{Variogram2015}. The lag tolerance $t$ gives us the maximum allowed deviation from $m\in\mathbb{R}^+$ when the distance between two data points is still considered to be $m$ meters (see Figure \ref{Fig:semivariogram}). For example if $m=10$ meters and $t=1$ meter, then the semivariogram value $\gamma_1(10)$ for a single data point $\textbf{d}_i$ is calculated from the set $\Gamma = \{\textbf{d}_j \in \mathcal{D} \;|\; e(\textbf{c}_j, \textbf{c}_i) \in [9, 11]\}$. In other words, the data points in set $\Gamma$ are considered to be $10$ meters away from $\textbf{d}_i$. This is rarely exactly the case and hence we have to use the lag tolerance $t$. The lag tolerance values in the experimental cases were selected to suite the resolution of the corresponding data. In the Moran's I autocorrelation plots we call \textit{baseline} the 0 correlation.

\subsection{CASE 1: Soil water permeability prediction based on soil type}
In this section we will consider the predictability of the soil water permeability levels based on the soil type. The response variable in this case is the water permeability exponent  value $y\in\mathbb{R}$, which is related both to the boreal soil type and to the water permeability itself. The exact relation between these two factors is presented in \citep{Pohjankukka2014b}. Optimal harvesting routes avoid areas with small water  permeability, where soil tends to stay moist and there is an elevated risk for ground damage and logistic problems. A reliable estimate of the water permeability distribution is needed when making routing decisions during the preliminary planning phase and during the harvest operations. The aim here is to increase the efficiency and minimize the harvesting costs.

\begin{figure}[t]
\begin{center}
\vspace{38pt}
\resizebox{\textwidth}{!}{%
\begin{minipage}{150mm}
\subfigure[]{
\resizebox*{7cm}{!}{\includegraphics{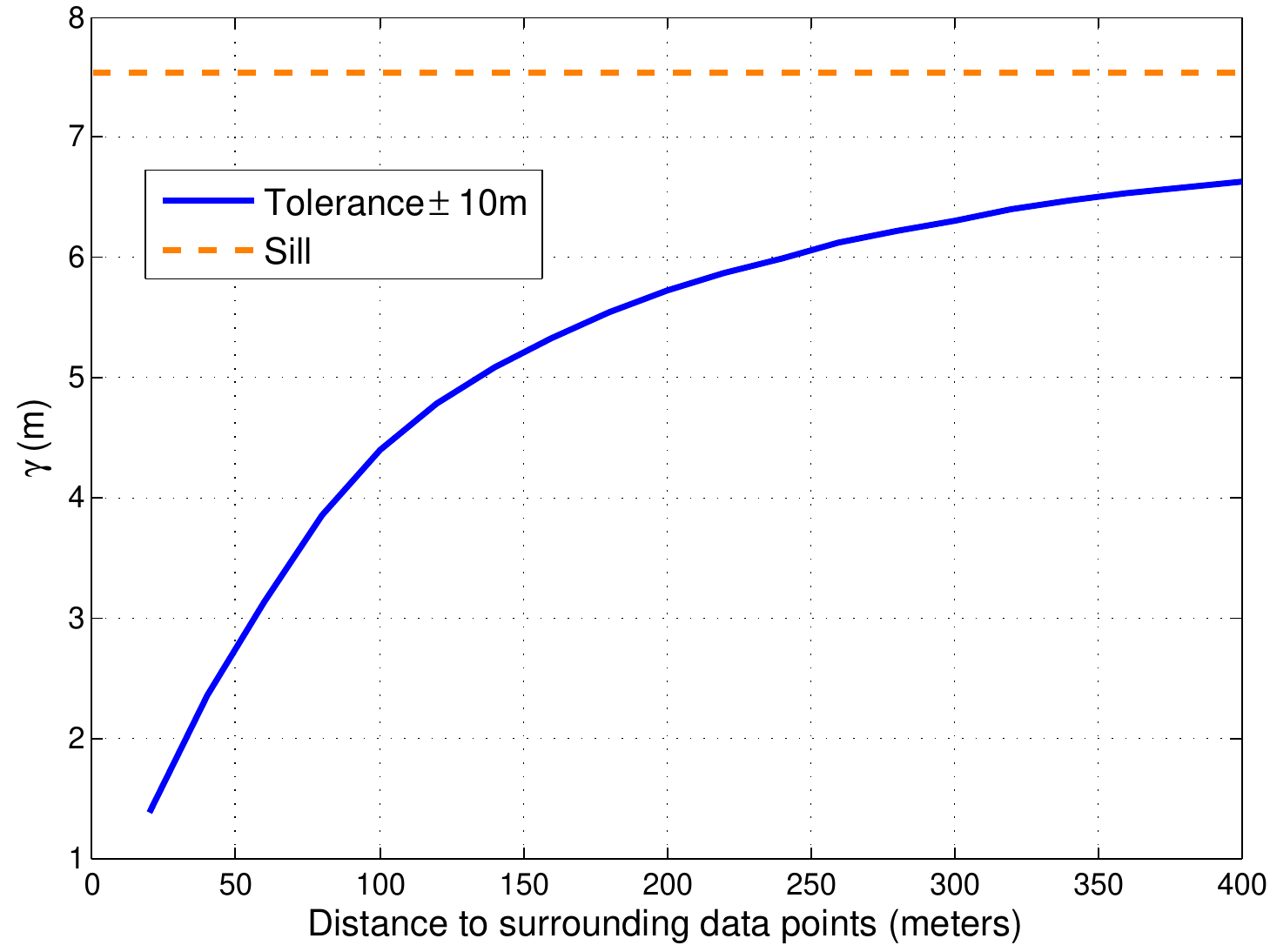}}}%
\subfigure[]{
\resizebox*{7cm}{!}{\includegraphics{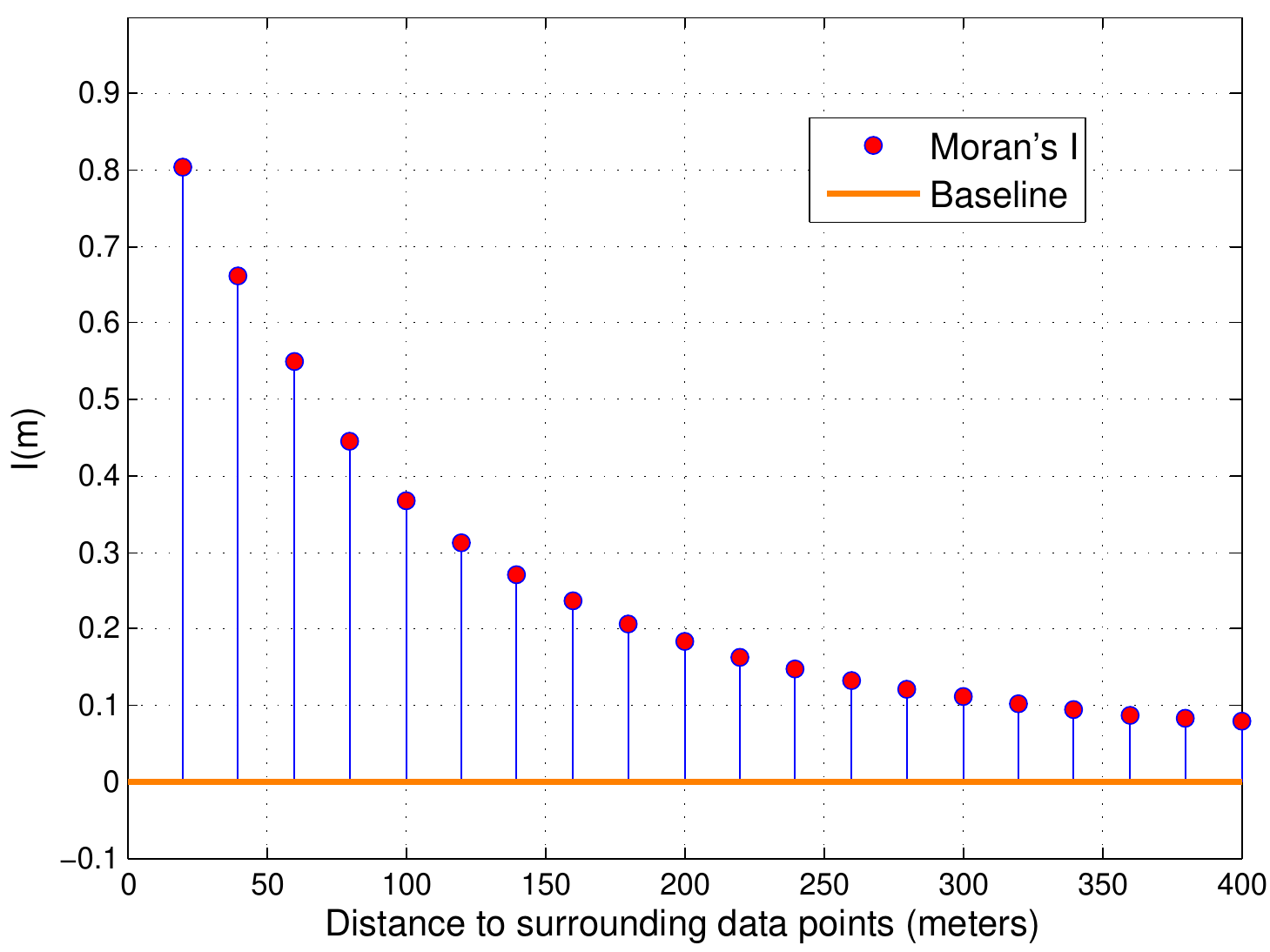}}}%
\caption{The semivariogram and Moran's I plot depicting the SAC of the water permeability exponent $y$ in case 1. (a) Semivariogram showing that $\gamma(m)$ stays below the sill with $t=10$ m. (b) Moran's I also revealing the presence of SAC in response value $y$.}%
\label{fig:RESPONSE_CASE1}
\end{minipage}
}
\end{center}
\end{figure}

The target area is located in the municipality of Parkano, which is a part of the Pirkanmaa region of Western Finland. The size of the target area is approximately $144\,\text{km}^2$ (ETRS-TM35FIN coordinates at 278 kmE, 6882 kmN, zone 35). When considering all the features including the derived ones, we had a total of 49 predictor features in the data set. In the analysis of case 1 a total of 361201 data points were available. A summary of the data sets is illustrated in Table \ref{table:case_data_sets} of section 3. In Figure \ref{fig:RESPONSE_CASE1} depicting the semivariogram and Moran's I plot for the water permeability exponent $y$ we can see a clear presence of SAC. The predicted water permeability exponent $\hat{y}_i$ for the $i$th data point $\textbf{d}_i=(\textbf{x}_i, y_{i}, \textbf{c}_i)$ using kNN-regression is defined as:
\setlength{\abovedisplayskip}{10pt}
\setlength{\belowdisplayskip}{10pt}
\begin{equation}
\label{predicted_response}
\hat{y}_i = \frac1k\sum_{y \in N_i}y,
\end{equation}
where $N_i$ is the set of water permeability exponent values $y$ of the $k$-nearest neighbors of $\textbf{d}_i$. 


\begin{figure}[t]
	\vspace{0pt}
	\centering
    \includegraphics[scale=0.55]{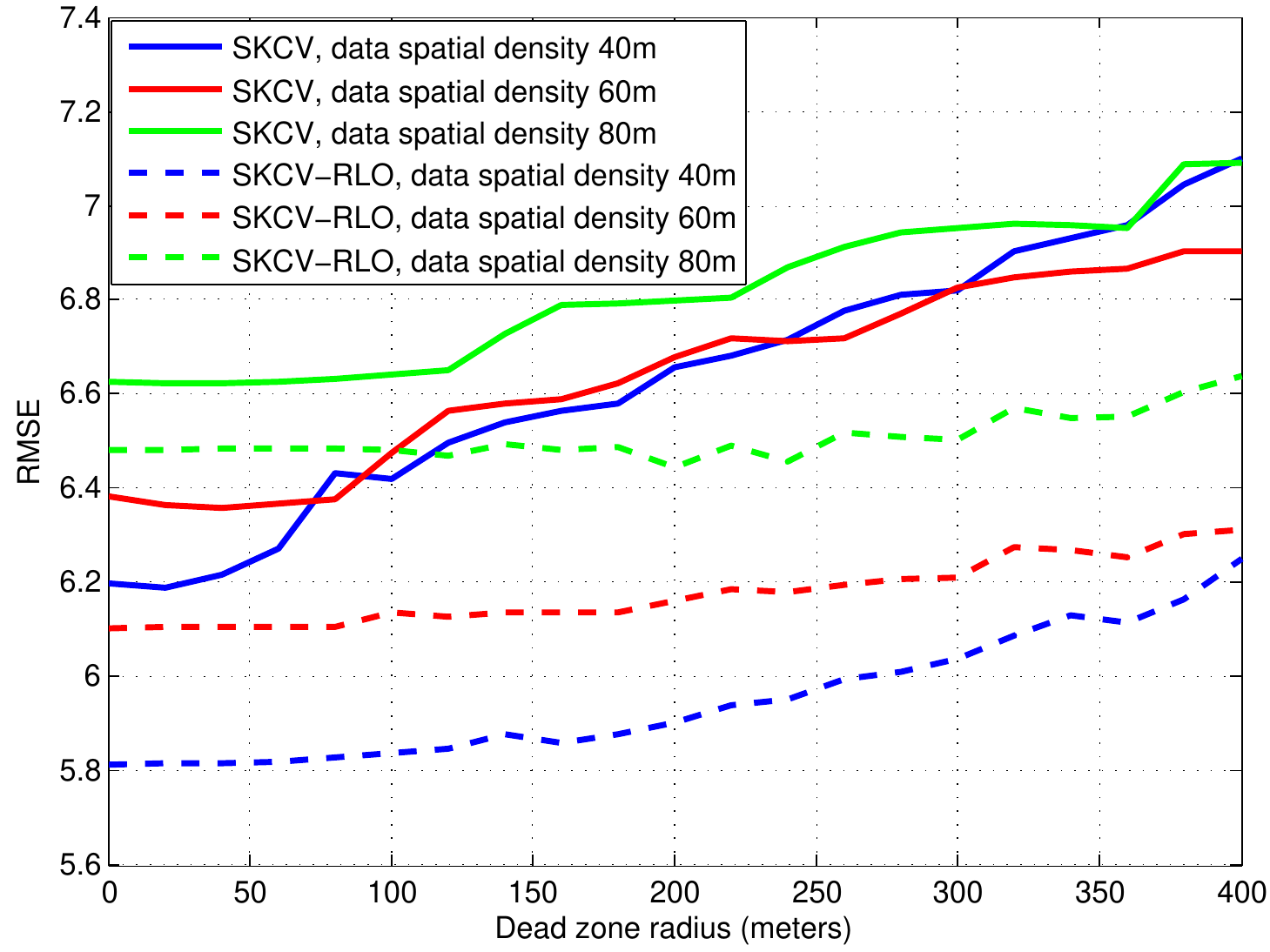}
    \vspace{2mm}
    \caption{Prediction performance estimates for 9NN using SKCV and SKCV-RLO in case 1. The curves are plotted with three spatial densities  to illustrate how the spatial density of the data set affects the results.}
    \label{fig::Parkano_results}
\end{figure}

The estimated prediction performance for 9-nearest neighbor (9NN) using SKCV is illustrated in Figure \ref{fig::Parkano_results}, which answers to research question (1) with various distance values $r_\delta$. The spatial density in the results describes how many data points are in a given space, i.e. it describes the sparsity of the data set. From Figure \ref{fig::Parkano_results} we notice a clear rise in the prediction error (RMSE) when the distance between prediction point and training data increases. This was an expected result based on the SAC discovered in the semivariogram and Moran's I plots in Figure \ref{fig:RESPONSE_CASE1}. With sparser data sets we notice the dead zone radius having a smaller effect on the results.

To measure how much the SKCV's performance decrease along the increasing dead zone radius is caused only by the decreased size of the training set, we implement additional analysis which we refer to as \emph{SKCV random-leave-out} (SKCV-RLO). SKCV-RLO is identical to the SKCV method (see Algorithm \ref{Algo} and Figure \ref{Fig:SKCV_illustration}) with the exception that instead of removing data points from the training set that are too close to the test data, i.e. inside the dead zone perimeter, we instead remove the same number of data points randomly from the training set as we would remove in SKCV. In Figure \ref{fig::Parkano_results} is illustrated the estimated prediction performance for 9NN using SKCV-RLO. On all spatial densities we notice SKCV-RLO being less sensitive to the number of data points removed from the training set giving more optimistic results than SKCV. This reinforces our claim that the prediction algorithm prefers to use data points which are geographically close to the prediction point and shows that random removal of training data points causes negligible change in prediction accuracy when compared with SAC-based data removal. 


\begin{figure}[t]
	\vspace{38pt}
	\centering
    \includegraphics[scale=0.25]{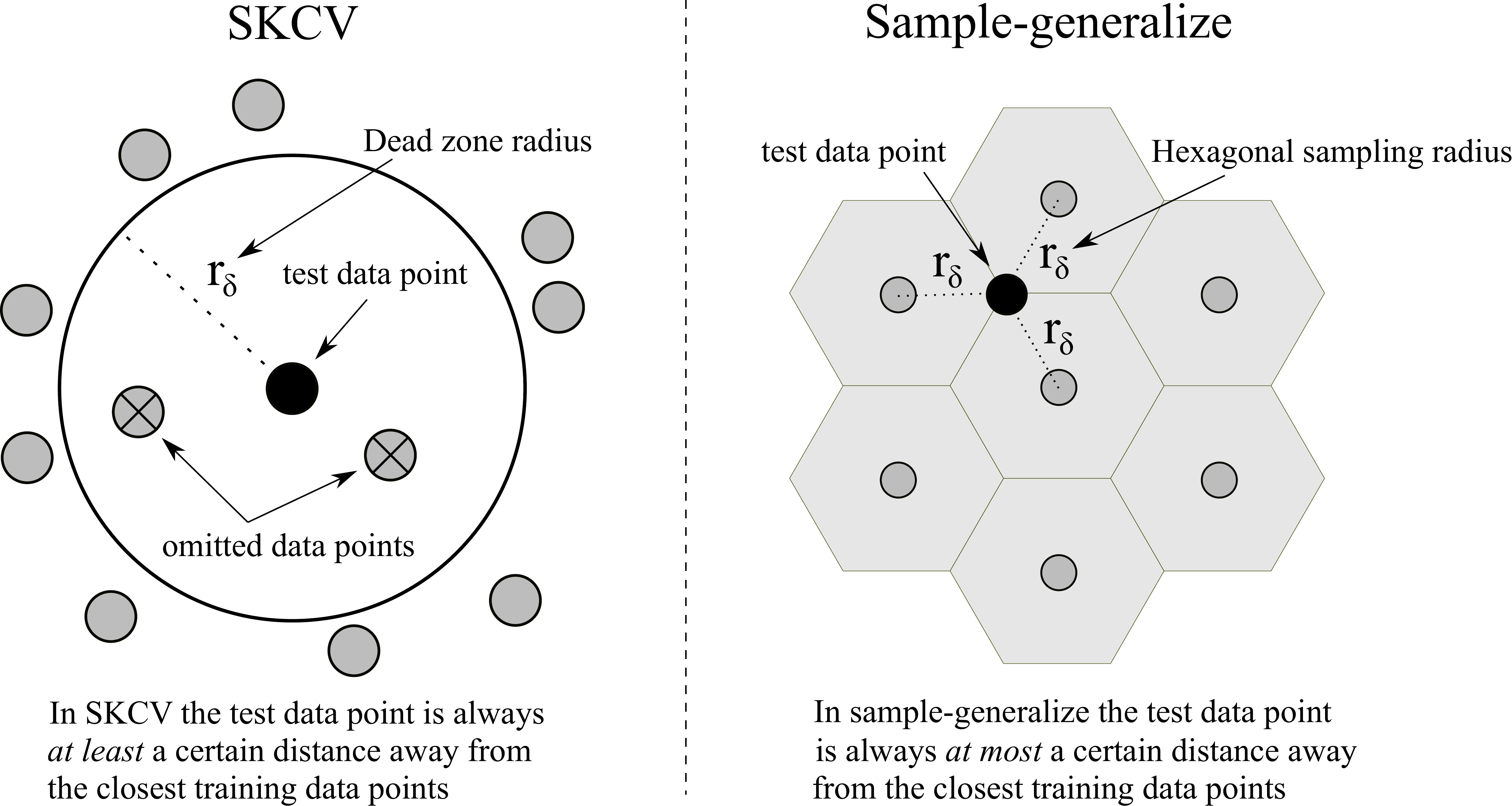}
    \vspace{2mm}
    \caption{\textit{Left:} In SKCV the test point is always at least $r_\delta$ meters away from training data. \textit{Right:} In sample-generalize procedure we sample data points (the gray points) using a hexagonal grid and predict the rest of the area around the sampled points. The black point represents a prediction point where the distance to training data is maximum, i.e. $r_\delta$ meters.}
    \label{Fig:SKCVSG}
\end{figure}

Next, we focus our attention on research question (2), i.e. how densely we should sample data points from a new research area to achieve a given prediction level. Imagine that there are two distinct geographical areas which we refer to as areas $A$ and $B$. In area $A$, there exists a data set of measurements gathered from a certain subset of its coordinates but there are no measurements from area $B$ yet. The aim is to perform a number of measurements from area $B$ in order to construct a model for predicting the rest of the measurement values for every possible point in area $B$. Performing measurements used to form a training set is expensive, and hence their number should be minimized under the constraint that at least a given prediction performance level is required. This trade-off between the number of training measurements and prediction performance is not known in advance and our hypothesis is that it can be estimated with SKCV on the existing data from area $A$. Namely, if the prediction performance estimate provided by SKCV with dead zone radius $r_\delta$ on area $A$ is as good or better than the required performance level, we hypothesize that we obtain as good prediction performance in area $B$ if we guarantee that the closest measurement points are at most at a distance of $r_\delta$ from every point in area $B$. Given this constraint, the number of measurement points in area $B$ is minimized via hexagonal sampling \citep[see e.g.][]{JMR58076}. To support our hypothesis (i.e. using SKCV to estimate the trade-off between number of measurement points and prediction performance) we use an auxiliary method called \textit{sample-generalize}. In the sample-generalize procedure we firstly sample training data points hexagonally (e.g. measure their response variables) with sampling radius $r_\delta$, and secondly we use this data to train a model for predicting the responses from the rest of the area. Right side of Figure \ref{Fig:SKCVSG} illustrates the sample-generalize procedure. Note that SKCV is inherently more pessimistic than sample-generalize since the prediction point is always \textit{at least} $r_\delta$ meters away from training data, whereas in sample-generalize the prediction point is always \textit{at most} $r_\delta$ meters away from training data (see Figure \ref{Fig:SKCVSG}). 

\begin{figure}[t]
\begin{center}
\vspace{38pt}
\resizebox{\textwidth}{!}{%
\begin{minipage}{150mm}
\subfigure[]{
\includegraphics[scale=0.3]{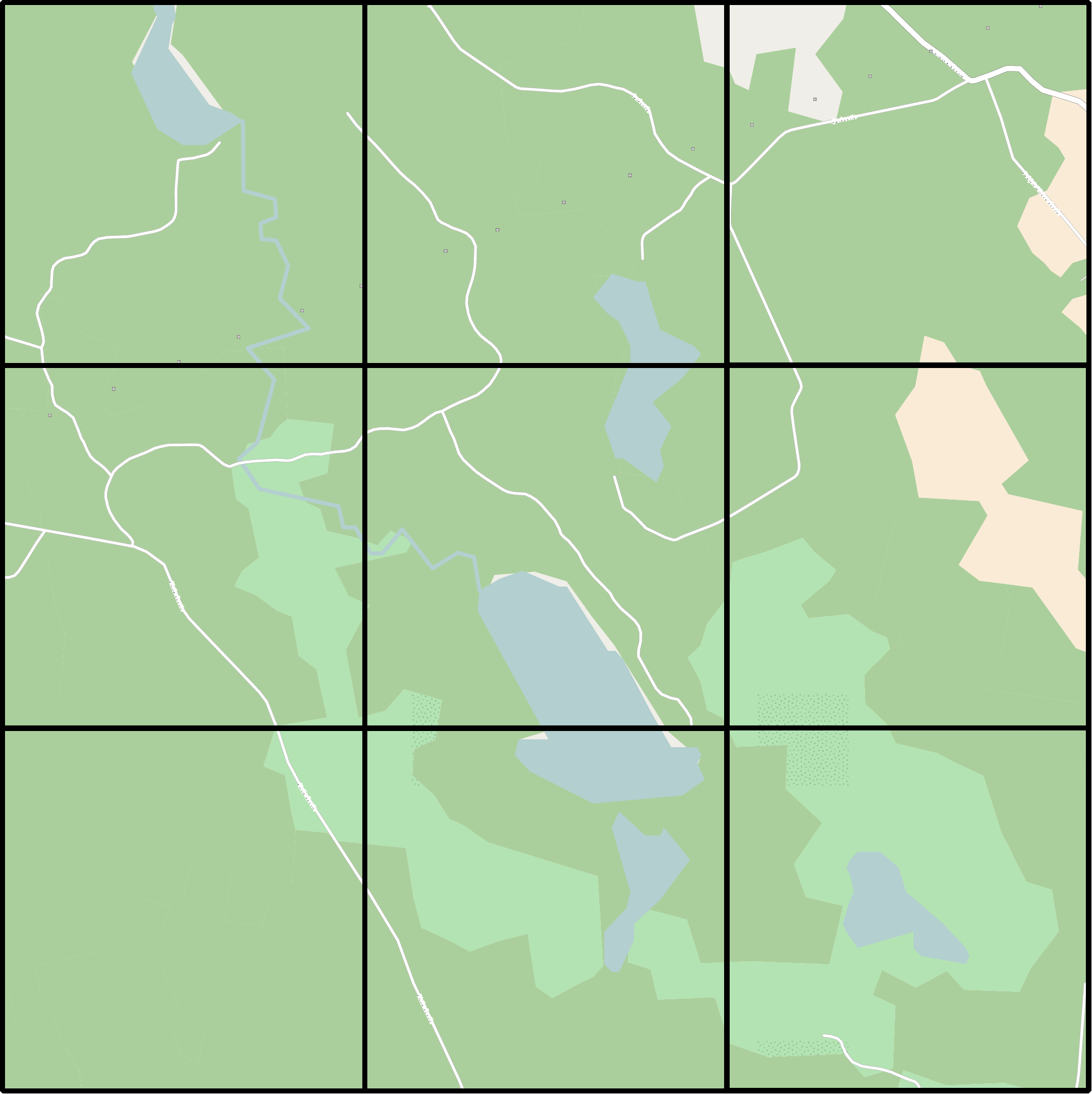}
\label{fig:3by3grid}
}%
\subfigure[]{
\includegraphics[scale=0.55]{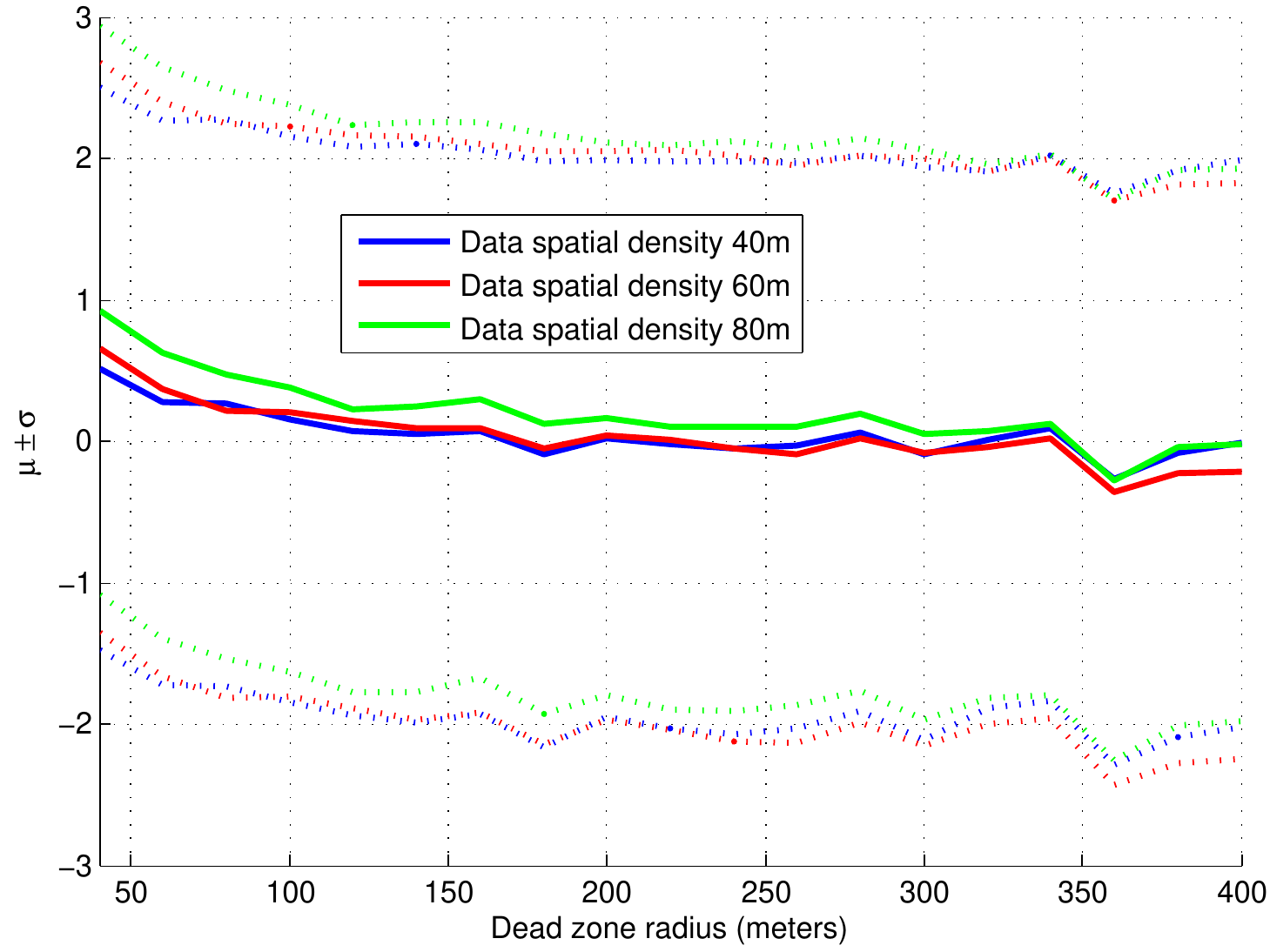}
\label{fig:SKCV_SG_DIFF_BIAS_VAR}
}%
\caption{(a) Division of a research area into nine smaller subareas using a 3x3 grid. Each smaller area is $16\,\text{km}^2$ in size and consists from approximately 40,000 data points. (b) Bias-variance $(\mu \pm \sigma)$ plot for the difference between the prediction performance estimate produced by SKCV and the actual prediction performance of sample-generalize of the 72 (9*8) area pairs. Solid curves represent the mean $\mu$ and dashed lines standard deviation $\sigma$. Different colors represent different spatial densities for the data set in area $A$ where SKCV is implemented.}

\end{minipage}
}
\end{center}
\end{figure}

In order to inspect the goodness of SKCV as an estimator of the prediction performance of sample-generalize we implement a bias-variance analysis for nine smaller subareas formed using a 3x3 grid in the Parkano research area (see Figure \ref{fig:3by3grid}). We do this by firstly forming 72 $(A, B)$ area pairs (from 3x3 grid we get 9*8=72 area pairs, i.e. each smaller area has 8 pair possibilities) from the nine smaller subareas. Secondly, for each of the area pairs $(A,B)$ we calculate the prediction performance estimate with SKCV on area $A$ ($result_A$) and the prediction performance of sample-generalize on area $B$ ($result_B$) and then we take the difference of them ($result_A-result_B$). Lastly, we calculate the mean and standard deviation of the  differences on the 72 area pairs. The resulting bias-variance plot is illustrated in Figure \ref{fig:SKCV_SG_DIFF_BIAS_VAR}. From the plot we see that the SKCV estimates tend to be pessimistically biased on the range $r_\delta \in [0, 150]$ meters. In range $r_\delta \in [150, 340]$ meters the SKCV estimation is almost unbiased and in range $r_\delta \in [340, 400]$ meters it is optimistically biased. The results are pretty stable on all spatial densities for SKCV, the spatial density seems to shift the results simply by a constant value. 


\subsection{CASE 2: Soil water permeability prediction based on field measurements}
\begin{figure}[t]
\begin{center}
\vspace{38pt}
\resizebox{\textwidth}{!}{%
\begin{minipage}{150mm}
\subfigure[]{
\resizebox*{6.5cm}{!}{\includegraphics{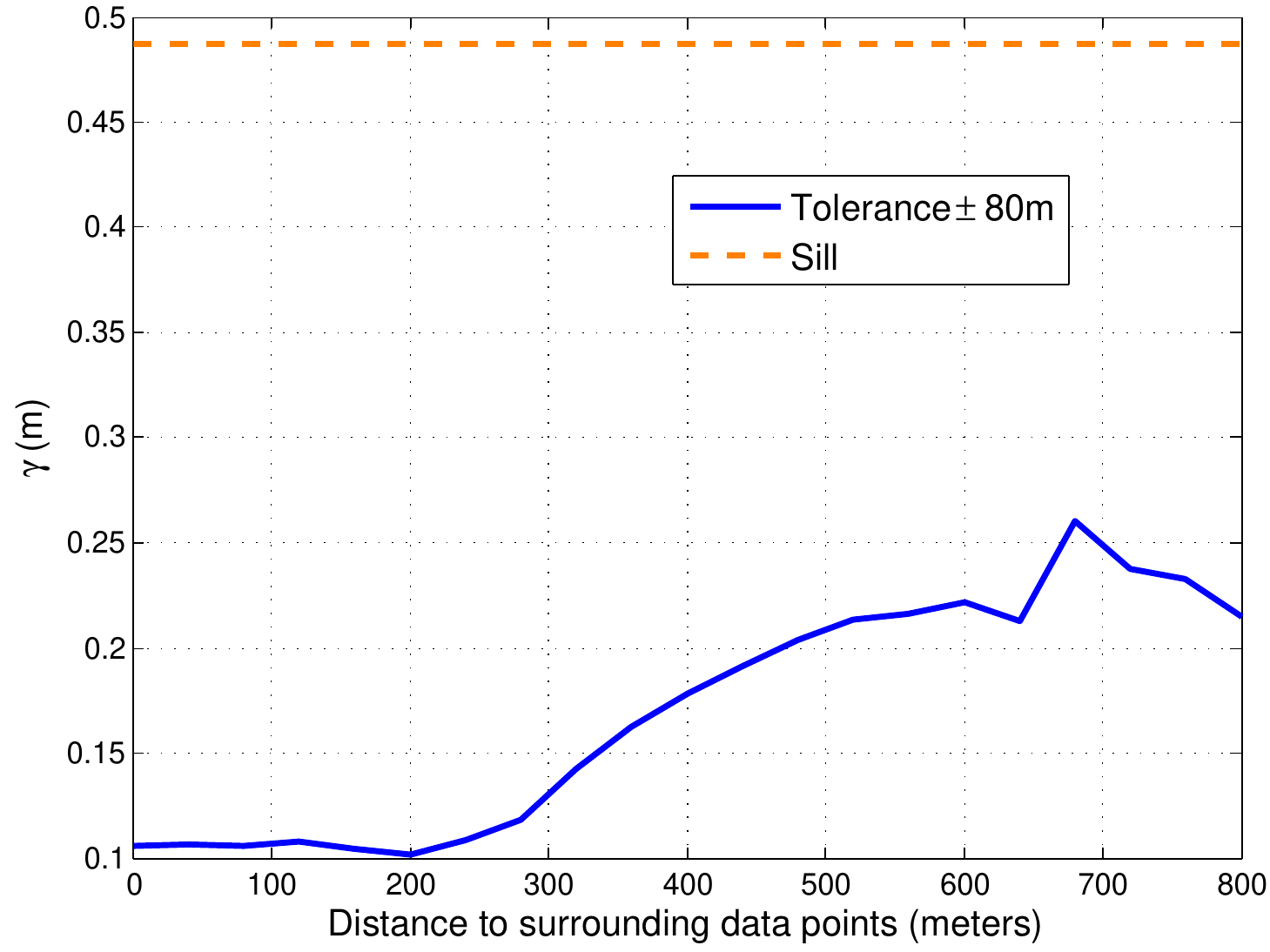}}}%
\subfigure[]{
\resizebox*{6.5cm}{!}{\includegraphics{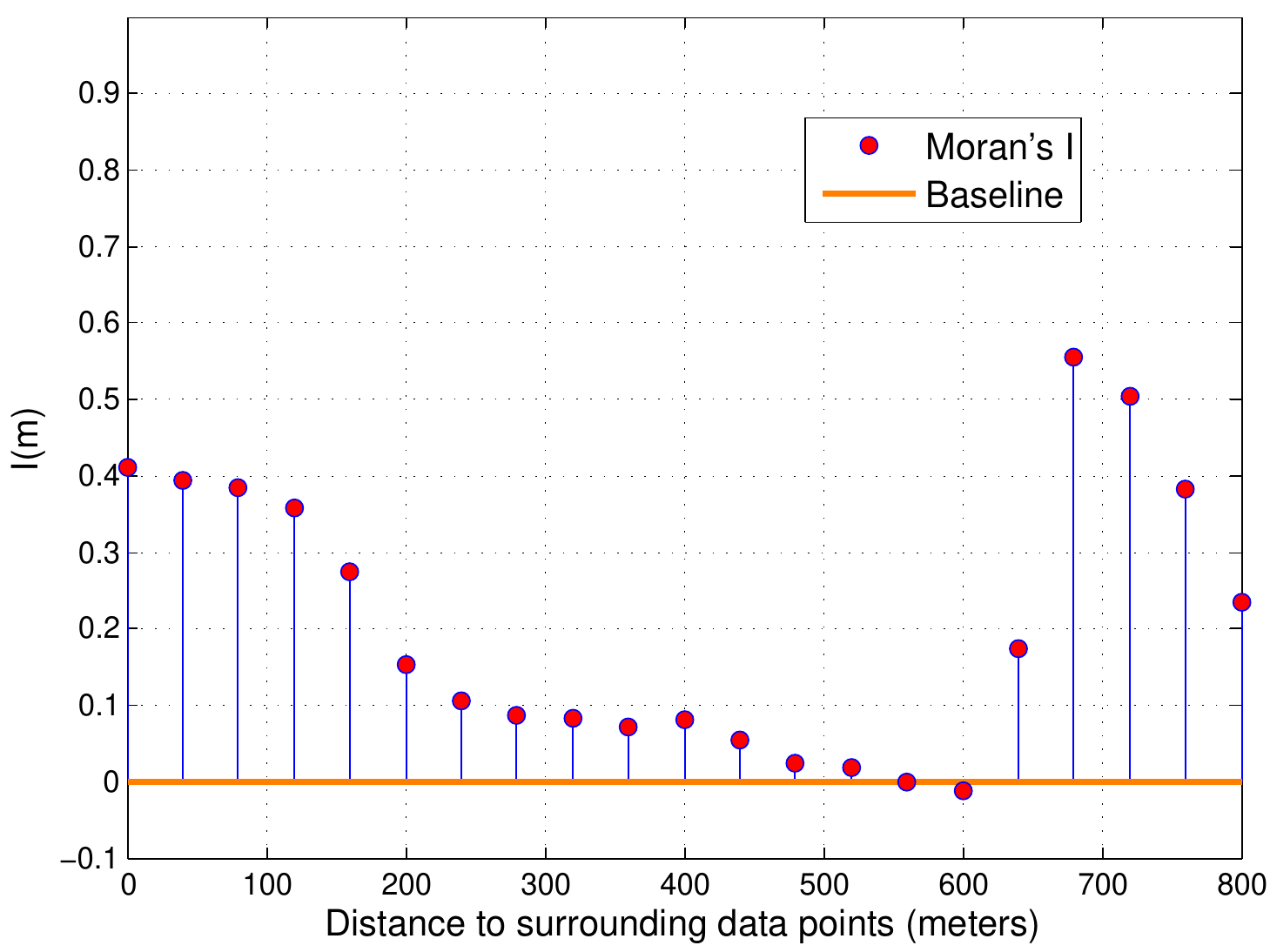}}}\\
\caption{The semivariogram and Moran's I plot depicting the SAC of the response value of case 2. (a) Semivariogram with $t=80$ m. (b) Moran's I.}
\label{fig:RESPONSE_CASE2}
\end{minipage}
}
\end{center}
\end{figure}

In this section we consider the predictability of forest soil water permeability based on field measurement data. The difference between the response variables in cases 1 and 2 is that in the case 1 the water permeability exponent $y$ is based on remote sensing data and in the case 2, $y$ is based on field measurements. Semivariogram and Moran's I plot for the response variable is presented in Figure \ref{fig:RESPONSE_CASE2} which show clear SAC in the data. There is more variability in the SAC of case 2 than case 1 but we must note that the data set in case 2 was much smaller and more sparse.


\begin{figure}[t]
\begin{center}
\vspace{0pt}
\resizebox{\textwidth}{!}{%
\begin{minipage}{150mm}
\centering
\includegraphics[scale=0.55]{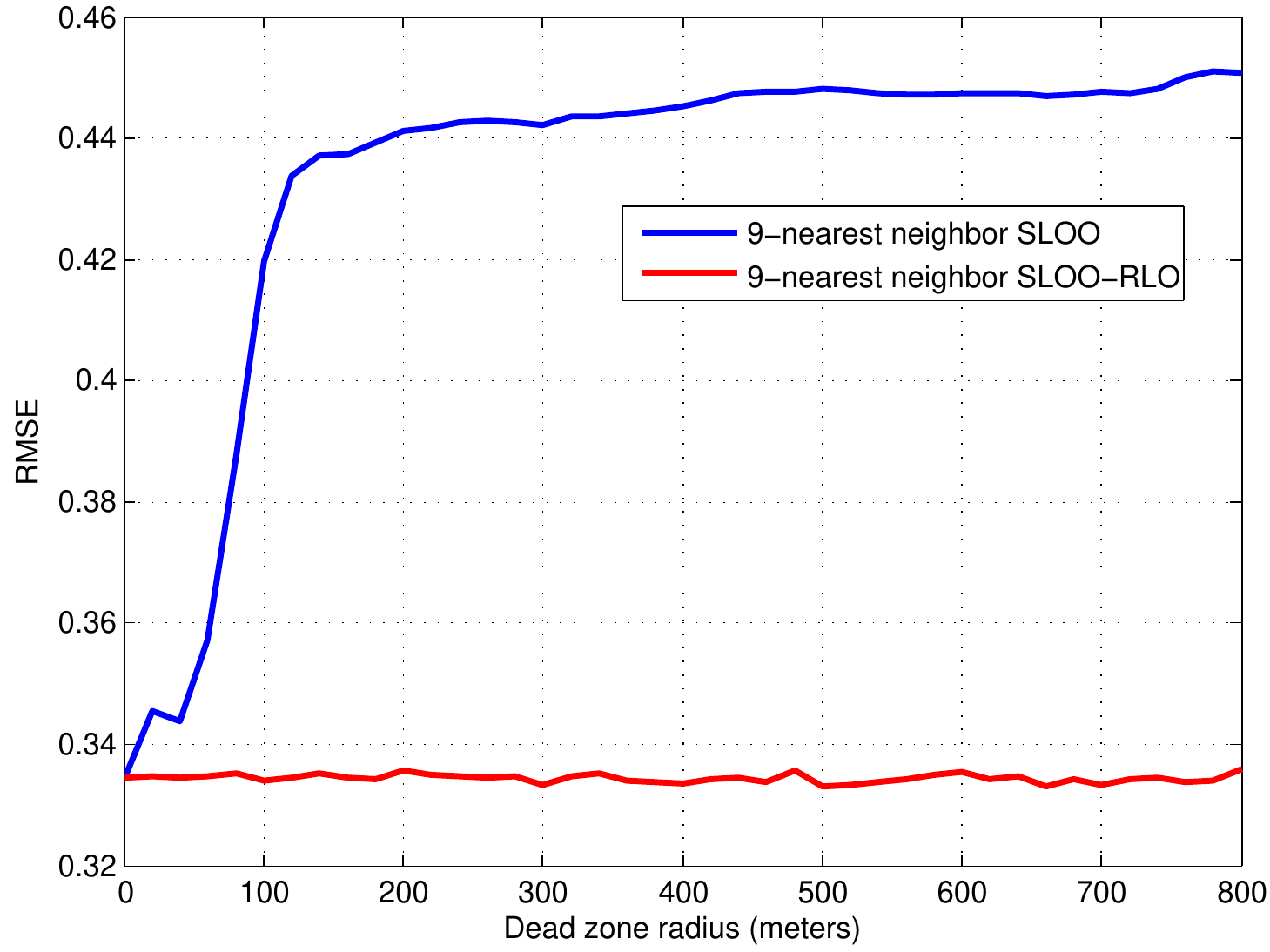}
\caption{The SLOO and SLOO-RLO results in the Pomokaira analysis. The y-axis corresponds to the RMSE and x-axis to the length of dead zone radius $r_\delta$.}
\label{fig:Pomokaira_results}
\end{minipage}
}
\end{center}
\end{figure}

The research area is located in Pomokaira, the northern part of the municipality of Sodankyl\"a, which is a part of Finnish Lapland. The size of the target area is $18432 \;\text{km}^2$. The center point of the rectangular target area is at ETRS-TM35FIN coordinates 7524 kmN, 488 kmE, zone 35. A total of 1691 data points were collected around the research area. The distances between the data points was much larger and they were not available from the entire research area when compared with the case 1 data set. 102 feature variables were used for predicting the response value i.e. the water permeability exponent $y$. The used data sets in case 2 are shown in Table \ref{table:case_data_sets}. The response variable $y$ is predicted in exactly the same way as in case 1 using kNN-regression in Equation \ref{Fig:harvester_illustration}.

Because the number of data points was significantly lower when compared with case 1 it was computationally feasible to implement SLOO and SLOO-RLO analyses on the data. The SLOO and SLOO-RLO results of case 2 are illustrated in Figure \ref{fig:Pomokaira_results}. The SLOO results show a clear drop in the prediction performance as the dead zone radius $r_\delta$ is increased. A high optimistic bias is observed from the SLOO-RLO results when compared with SLOO. The SLOO results indicate that the prediction performance decreases radically after the distance between test and training data is approximately 40-50 meters. The effect of SAC can clearly be noted in these results.

\subsection{CASE 3: Soil track damage classification} 
\begin{figure}[b]
\begin{center}
\vspace{38pt}
\resizebox{\textwidth}{!}{%
\begin{minipage}{150mm}
\subfigure[]{
\resizebox*{6.5cm}{!}{\includegraphics{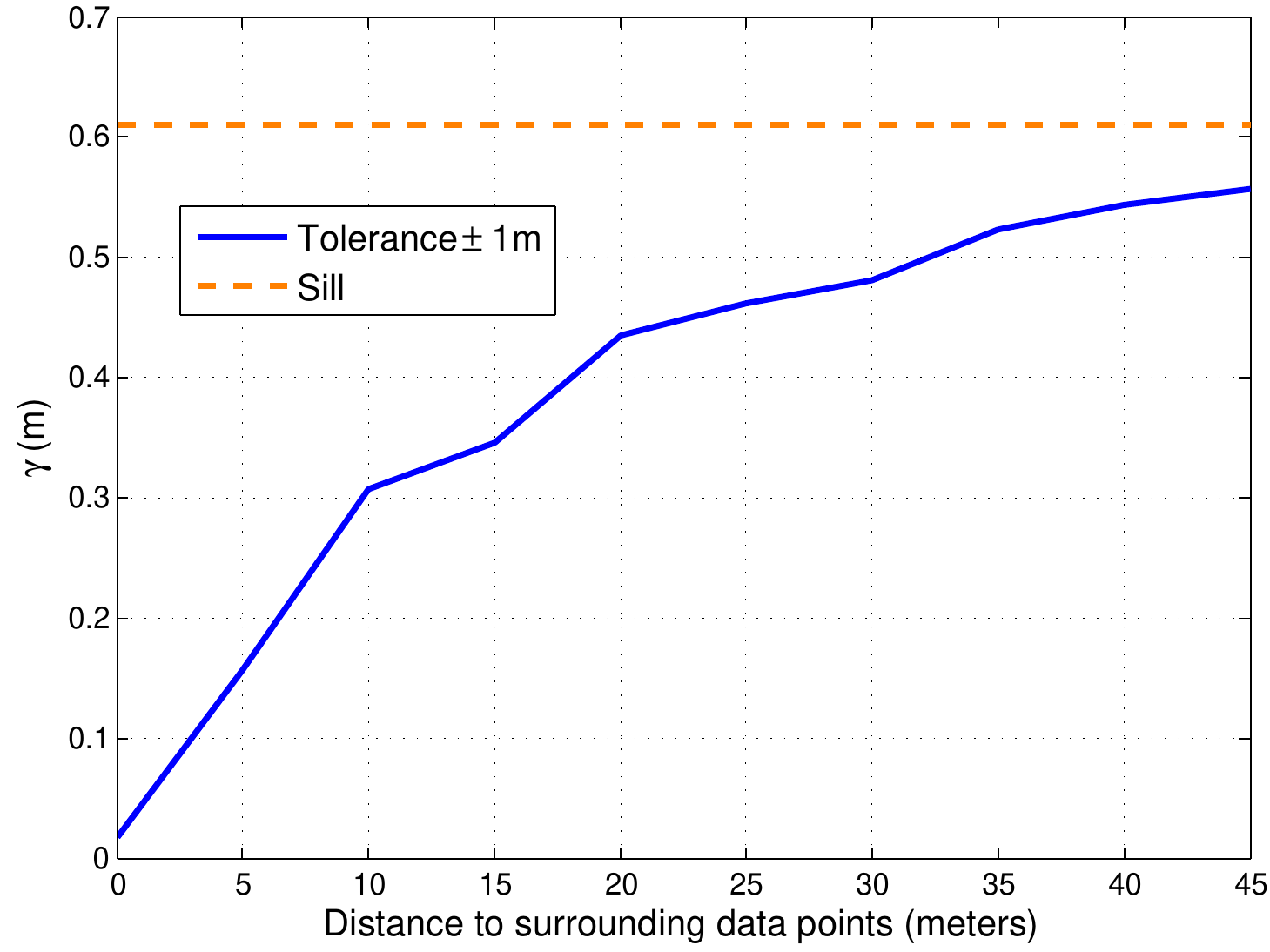}}}%
\subfigure[]{
\resizebox*{6.5cm}{!}{\includegraphics{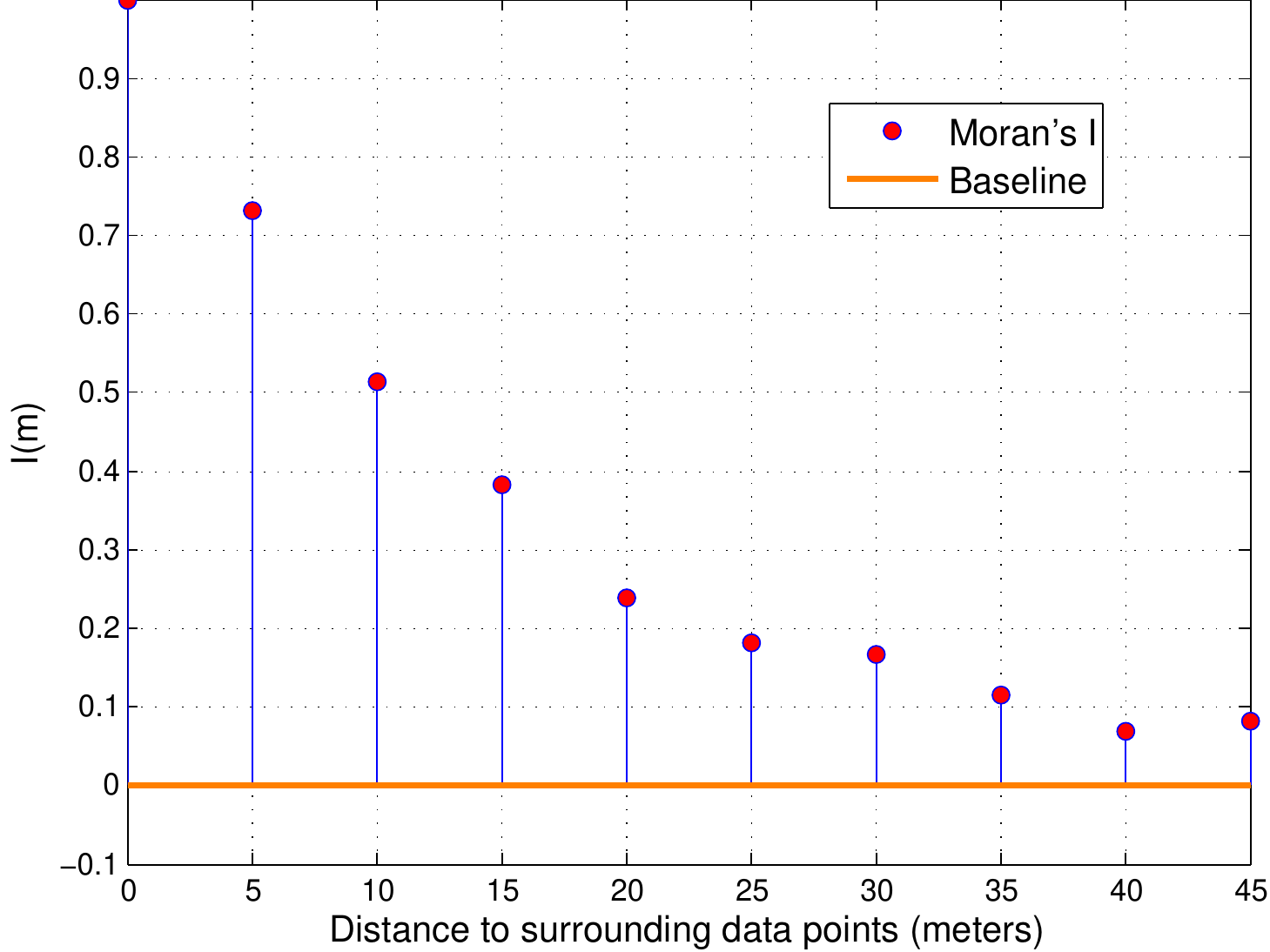}}}\\
\caption{The semivariogram and Moran's I plot depicting the SAC of the response value of case 3. (a) Semivariogram with $t=1$ m. (b) Moran's I.}
\label{fig:RESPONSE_CASE3}
\end{minipage}
}
\end{center}
\end{figure}

In this case the goal is to assess the classification of forest harvester track damage. In other words, the task is to predict the damage that would occur to a soil point if a forest harvester drives through it. In particular, damage means the depression caused on the soil by the harvester. Track damage is affected by soil type, humidity, penetration resistance etc. The penetration resistance of soil is an important factor in forest harvesting operations which must be accounted for in order to prevent additional costs for harvesting. Peat areas for example cause challenging soil conditions for heavy machinery and extra carefulness is needed there. It is both expensive and laborious operations to get sunken forest harvesters out from peats. Therefore it is important to select harvesting routes which have the highest possible penetration resistance. As in cases 1 and 2, the semivariogram and Moran's I plot for the response variable of case 3 are presented in Figure \ref{fig:RESPONSE_CASE3}, which also show a clear presence of SAC. Note that the track damage is an ordinal variable consisting from three classes and hence it was also possible to construct a variogram in this case.

The research area consists from 13 different harvesting areas in Pieks{\"a}m{\"a}ki, a municipality located in the province of Eastern Finland 62$^{\circ}$18'N 27$^{\circ}$08'E. A total of 83 feature variables were used for classifying the soil damage. The sizes of the data sets collected from each of these areas ranged from hundreds of samples to thousands of samples. The total number of data points was 11795. As in cases 1 and 2 the used data sets in case 3 are shown in Table \ref{table:case_data_sets}. In case 3 the predicted response value of $\hat{y}_i$ (track damage class) is defined as the mode of set $N_i$ (kNN-classification), where $N_i$ is again the set of k-nearest neighbors of data point $\textbf{d}_i$.


\begin{figure}[t]
\begin{center}
\vspace{38pt}
\resizebox{\textwidth}{!}{%
\begin{minipage}{150mm}
\centering
\includegraphics[scale=0.55]{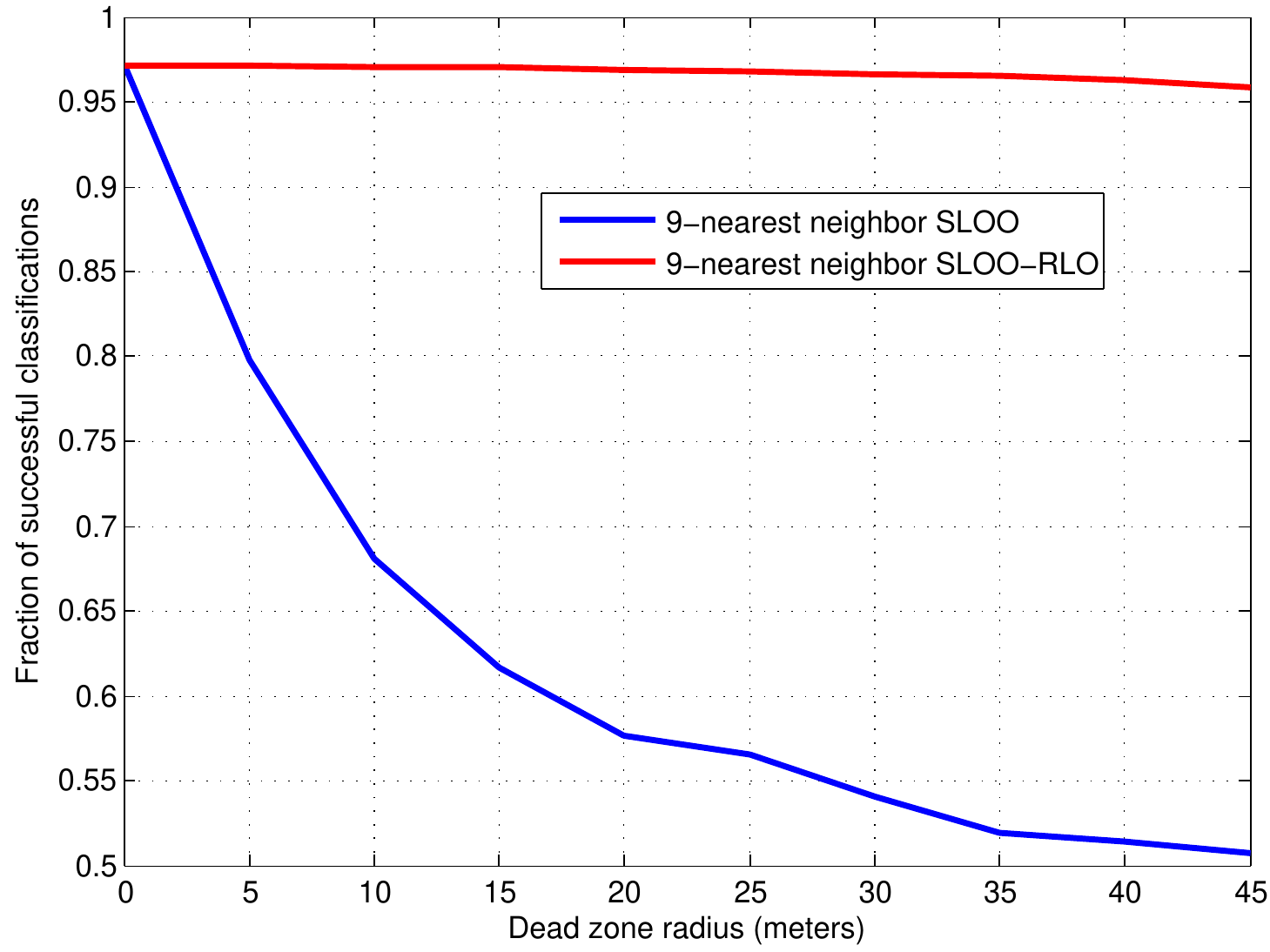}
\caption{The SLOO and SLOO-RLO results in the Pieks\"am\"aki analysis. The y-axis corresponds to the fraction of successful classifications and x-axis to the length of dead zone radius $r_\delta$.}
\label{fig:Pieksamaki_results}
\end{minipage}
}
\end{center}
\end{figure}

The SLOO and SLOO-RLO analyses were conducted on each of the 13 harvest areas separately because the distances between the harvest areas were in worst cases dozens of kilometers. On each of these areas the SLOO and SLOO-RLO procedures were implemented and the results were averaged over all areas. Figure \ref{fig:Pieksamaki_results} presents the SLOO and SLOO-RLO results for case 3. Similarly as in cases 1 and 2, the results in case 3 confirm the effect of SAC on prediction performance estimates. One can notice an exponential form decay in the SLOO results as a function of dead zone radius $r_\delta$ whereas the SLOO-RLO results are almost unchanged as it was also in case 2. In the worst case we have approximately 40\% difference in the results between SLOO and SLOO-RLO. 

\section{Conclusion}
Spatio-temporal autocorrelation is always present with GIS-based data sets and needs to be accounted for in machine learning approaches. As discussed above, traditional model performance criteria such as the CV method omit the consideration of the effect of SAC in the performance estimations with natural data sets. To account for the SAC in GIS-based data sets we demonstrated by the means of three experiments that the SKCV method can be used for estimating the prediction performance of spatial models without the optimistic bias due to SAC, while the ordinary CV can cause highly optimistically biased prediction performance estimates. We also showed that SKCV can be used as a data sampling density selection criterion for new research areas, which will result in reduced costs for data collection.

\section*{Acknowledgements}
We want to thank the Natural Resources Institute Finland (LUKE), Geological Survey of Finland (GTK), Natural Land Survey of Finland (NLS) and Finnish Meteorological Institute (FMI) for providing the data sets. This work was supported by the funding from the Academy of Finland (Grant 295336). The preprocessing of the data was partially funded by the Finnish Funding Agency for Innovation (Tekes).

\bibliographystyle{tGIS}
\bibliography{tGISguide.bib}

\end{document}